\documentclass[lettersize,journal]{IEEEtran}

\usepackage{amssymb,amsmath}
\usepackage{cite}
\usepackage{graphicx}
\usepackage{subfig}

\usepackage{psfrag}
\usepackage{url}
\usepackage[latin1]{inputenc}
\usepackage[absolute,overlay]{textpos}
\usepackage{gensymb}
\usepackage{cases}
\usepackage[font=footnotesize]{caption}
\usepackage{float}
\usepackage[linesnumbered,ruled,lined]{algorithm2e}
\newcommand{\argmax}[1]{{\operatorname{arg}\,\max_{#1}}\,}

\usepackage{pgf}

\usepackage[nocomma]{optidef}

\usepackage{amsthm}

\usepackage{booktabs}
\usepackage{xcolor,soul}

\usepackage{textcomp}
\usepackage{bbm}
\def\BibTeX{{\rm B\kern-.05em{\sc i\kern-.025em b}\kern-.08em
    T\kern-.1667em\lower.7ex\hbox{E}\kern-.125emX}}

\usepackage{tabularx}
\usepackage{makecell}
\usepackage{multirow}

\newcommand{\sfrac}[2]{\,^{#1}\!/\!_{#2}}

\begin{document}

\title{
An Offline Multi-Agent Reinforcement Learning Framework for Radio Resource Management 
}
\author{
	\IEEEauthorblockN{Eslam Eldeeb and Hirley Alves
}
	
    \thanks{Eslam Eldeeb and Hirley Alves are with the Centre for Wireless Communications (CWC), University of Oulu, Finland. (e-mail: eslam.eldeeb@oulu.fi; hirley.alves@oulu.fi).
    }
    
    \thanks{This work was supported by 6G Flagship (Grant Number 369116) funded by the Research Council of Finland.}
}
\maketitle

\begin{abstract}

Offline multi-agent reinforcement learning (MARL) addresses key limitations of online MARL, such as safety concerns, expensive data collection, extended training intervals, and high signaling overhead caused by online interactions with the environment. In this work, we propose an offline MARL algorithm for radio resource management (RRM), focusing on optimizing scheduling policies for multiple access points (APs) to jointly maximize the sum and tail rates of user equipment (UEs). We evaluate three training paradigms: centralized, independent, and centralized training with decentralized execution (CTDE). Our simulation results demonstrate that the proposed offline MARL framework outperforms conventional baseline approaches, achieving over a $15\%$ improvement in a weighted combination of sum and tail rates. Additionally, the CTDE framework strikes an effective balance, reducing the computational complexity of centralized methods while addressing the inefficiencies of independent training. These results underscore the potential of offline MARL to deliver scalable, robust, and efficient solutions for resource management in dynamic wireless networks.
\end{abstract}

\section{Introduction}\label{sec:introduction}

The road toward future intelligent wireless communication systems, such as the one envisioned by 6G, is paved with a growing interest in applying machine learning / artificial intelligence (ML/AI) to wireless systems~\cite{8714026,mahmood2021machine}. 
Machine learning and artificial intelligence (ML/AI) techniques have been instrumental in advancing beyond 5G systems. They are poised to play a more critical role in developing 6G networks. Given the heightened scale, complexity, and distributed nature of 6G, these technologies are essential for effectively addressing such intricate challenges ~\cite{10198239,9845353}. 
These challenges include (but are not limited to) radio resource management (RRM), which is often too complex to be modeled using traditional statistical methods. The RRM problem is generally a non-convex optimization problem, and its complexity increases tremendously as the network grows.

In the literature, RRM has been addressed using information theory~\cite{7421354}, geometric programming~\cite{4600228}, and game theory~\cite{6845058}. However, these algorithms may fail due to the dynamic behavior of the wireless systems. In this regard, online reinforcement learning (RL) has shown a promising contribution towards solving RRM problems~\cite{10004901}. Online RL involves an agent that interacts with the environment, observes its condition (state), takes a decision (action) and receives a feedback signal (reward) indicating the quality of the decision. Online RL techniques are suitable for RRM as they can solve the complex RRM problem in a model-free manner, without deployment knowledge. In addition, it benefits from the control and feedback methods in the wireless network to iteratively optimize and update the designed algorithm.

Recent advances in Online RL have witnessed the rise of two well-established RL frameworks: deep RL and multi-agent reinforcement learning (MARL). Deep RL combines robust deep neural networks (DNNs) with RL~\cite{mnih2015human}. This eases optimizing complex and large-scale environments. On the other hand, MARL enables joint decision-making optimization (policy optimization) of multiple agents~\cite{da2024distributed}. MARL algorithms vary from cooperative MARL~\cite{oroojlooy2023review}, where various agents cooperate towards one goal, and competitive MARL~\cite{tampuu2017multiagent}, where multiple agents compete against each other. We focus on cooperative MARL as we aim to optimize the scheduling policy of several entities to achieve a joint goal in the system. The cooperative MARL problem itself varies according to the agents' communication rate~\cite{eldeeb2023traffic}. For instance, decentralized solutions assume no communication between the agents, whereas centralized techniques allow complete communication between the agents~\cite{marl-book}. Recent techniques propose mixed centralized and decentralized techniques~\cite{sunehag2017valuedecompositionnetworkscooperativemultiagent}.

Online MARL faces significant challenges when deployed to real-time wireless scenarios. First, it relies on online interaction with the environment to explore and visit the environment states. This online interaction might not be feasible, safe, timely, or costly. Second, some MARL variants, such as centralized MARL methods, enable interaction between the agents, which adds an extra layer of complexity and overhead to the environment. These problems can addressed by optimizing the policy offline via a static dataset pre-collected using a behavioral policy. Thus, undesirable online interactions are mitigated. This opens the door for offline MARL.

\subsection{Offline MARL}

\emph{Offline MARL} considers an offline static dataset to be used to optimize the policies of multiple agents~\cite{levine2020offline}. Offline MARL assumes that the agents can not interact with the environment during the optimization (training) phase. After training, the agents deploy the policies they have learned online. The offline dataset is usually collected using a behavioral policy, which is a policy designed using known traditional methods or even randomly. Offline MARL overcomes the safety and cost problems accompanied by online MARL by mitigating online interaction. Moreover, it limits signaling overhead and complex communication requirements between the agents by moving policy optimization offline. In addition, since the training is performed offline, it can be easily transferred to a powerful central unit, removing computational burdens from the limited resources of wireless entities.

Adapting traditional online MARL techniques to an offline setting introduces a distributional shift between the behavior and learned policies. This shift motivates overestimation of the unseen experiences in the dataset, uncertain policies, and training degradation~\cite{kostrikov2021offlinereinforcementlearningimplicit}. Several methods have suggested constraints on the difference between the behavior and learning policies, called behavior-constrained methods~\cite{schulman2015trust}. Another family of methods penalizes the value of out-of-distribution (OOD) actions (unseen actions in the dataset), which are called conservative methods~\cite{kumar2020conservative}. Conservative Q-learning (CQL) is a conservative offline RL technique that uses KL-divergence as a regularization parameter to penalize the weights of OOD actions. This work proposes an offline MARL algorithm based on CQL for the RRM problem.

\subsection{Related Work}
Over the past few years, many works in the literature have contributed to solving the RRM problem, mostly in an online fashion, for single and multi-agent reinforcement learning. Among the first to work on this problem is~\cite{6542770}, which proposes a (single-agent) reinforcement learning approach for self-organizing networks (SONs) in small cells. Then, \cite{8352517} proposes a deep RL algorithm for the spectrum sharing and resource allocation problem in cognitive radio systems. They adopt a deep RL algorithm for efficient power control so that the secondary user can share a common spectrum with the primary user. The authors in~\cite{eldeeb2022multi} propose an efficient resource and power optimization using reinforcement learning to jointly minimize the age-of-information (AoI) and transmission power of IoT sensors in unmanned aerial vehicles (UAVs) networks. In contrast, the work in~\cite{8931561} combines generative adversarial network (GAN) with deep RL for resource management and network slicing. A recent work in~\cite{10066838} solves the RRM problem using graph neural networks (GNNs). The authors formulate the problem as an unsupervised primal-dual problem. They develop a GNN architecture that parameterizes the RRM policies as a graph topology derived from the instantaneous channel conditions.

Several works have formulated MARL algorithms for the RRM problem and wireless communication. For example, the authors in~\cite{10293964} formulate the MARL problem for resource management in UAV networks. They present a comprehensive comparison between different MARL schemes. The authors in~\cite{naderializadeh2021resource} propose an online MARL algorithm for the RRM problem to maximize both sum and tail rates. In~\cite{8792117}, a dynamic power allocation is performed using MARL, where local observations are shared between nearby transmitters and receivers. The authors in~\cite{9217951} propose a distributed MARL approach for multi-cell wireless-powered communication networks to charge limited power users for efficient data collection wirelessly. In~\cite{9904510}, the authors addressed the dynamic resource management in X-subnetworks, where X refers to any entity such as a robot, vehicle, or module, and subnetworks refer to cells that can be part of a larger infrastructure. They propose combining MARL algorithms and attention-based layers to solve the resource management problem.

Even though offline RL and offline MARL are promising techniques, they have only recently begun to capture significant attention from the wireless communications community, e.g., \cite{eldeeb2024conservative, 10529190}.
In~\cite{eldeeb2024conservative}, the authors propose an offline and distributional MARL algorithm for resource management in UAV networks. The work in~\cite{10529190} proposes a single-agent offline RL algorithm for the RRM problem. It has proved that a mixture of datasets of multiple behavioral policies can lead to an optimal scheduling policy. However, they assume all access points can be modeled as one agent, thus neglecting the multi-agent scenario. The work in~\cite{eldeeb2024offlinedistributionalreinforcementlearning} proposes an offline and distributional RL algorithm for the RRM problem that combines deep RL with distributional RL offline to overcome the uncertainties of the wireless environment. Similarly, they only focus on the single-agent case. 

Most of the literature above suffers from significant drawbacks. First, the majority of these works targeted optimizing a single objective. However, the RRM problem targets multiple objectives, \textit{e.g.}, maximizing both sum and tail rates or AoI and pilot length. Second, some of these works considered the single-agent scenario and combined all agents in a centralized fashion. This is a notable concern as the network usually consists of many transmitters and users. Therefore, handling the RRM problem in a centralized fashion explodes the dimension and complexity of the RL problem, making it vulnerable to degrading performance. Finally, most of the existing works in the literature considered online RL or online MARL, which is unsafe, impractical, and very complex~\cite{meng2023offline} due to the need for a massive online interaction with the environment, especially in the multi-agent case. These challenges heavily affect the communication network due to the need for continuous communication between the agents, leading to significant signaling overhead.

\subsection{Main Contributions}

This work presents an offline MARL algorithm for RRM. We assume a general model and pose the RRM problem as a partially observable Markov decision process (MDP).
Offline MARL proposes multi-agent optimization using only an offline static dataset without any interaction with the environment. Hence, it fits the RRM problem where multiple agents cooperate to serve the users. 
To illustrate our results, but without loss of generality, we model our RRM problem as a joint optimization problem that includes sum and tail rates. We aim to reach a resource management policy that maximizes a linear combination of sum and tail rates. The main contributions of this paper are summarized as follows. 
\begin{itemize}
    \item We formulate the RRM problem using a partially observable Markov decision process (MDP). In addition, we present a preliminary result using online MARL.
    
    \item We propose two offline MARL algorithms: soft actor-critic (SAC), and conservative Q-learning (CQL). We present three variants of these offline MARL schemes using centralized learning, independent learning, and centralized training decentralized execution, respectively.
    
    \item We compare the three offline MARL schemes to four benchmarks from the literature. The proposed MARL schemes outperform the baseline models regarding both sum and tail rates. 
    
    \item We demonstrate that centralized training decentralized execution approaches overcome the complexity of centralized training MARL and the inefficiency of the independent training MARL. Our algorithm surpasses existing schemes by more than $50 \%$ gain regarding the linear combination of sum and tail rates.
\end{itemize}

The rest of the paper is organized as follows: Section~\ref{sec:system_model} introduces the RRM system model. The MARL formulation is proposed in Section~\ref{sec:MARL}. Section~\ref{sec:OMARL} depicts the proposed offline MARL algorithm. Simulation analysis is presented in Section~\ref{sec:results}, and the paper is concluded in Section~\ref{sec:conclusions}. Table~\ref{Abbrev} presents the list of abbreviations, while Table~\ref{Notations} summarizes the list of symbols and notations.

\begin{table}[t!]
\centering
\caption{List of abbreviations.}
\label{Abbrev}
\begin{tabular}{p{.2\columnwidth}|p{.7\columnwidth}}
\hline
\textbf{Abbreviation} & \textbf{Description} \\ \hline 
AI & Artificial intelligent \\ 
AoI & Age-of-information \\ 
AWGN & additive white Gaussian noise \\ 
BCQ & Behavior constrained Q-learning \\ 
CDF & Cumulative distribution function \\ 
CQL & Conservative Q-learning \\ 
AP & Access point \\ 
C-MARL & Centralized multi-agent reinforcement learning \\ 
CTDE & Centralized training and decentralized execution \\ 
DNN & Deep neural network \\ 
DQN & Deep Q-network \\ 
DRL & Distributional reinforcement learning \\ 
GAN & Generative adversarial network \\ 
GNN & Graph neural network \\ 
I-MARL & Independent training MARL \\ 
ITLinQ & Information-theoretic link scheduling \\ 
MARL & Multi-agent reinforcement learning \\ 
MDP & Markov decision process \\ 
OOD & Out-of-distribution \\ 
PF & Proportional fairness \\ 
PO-MDP & Partially-observable Markov decision process \\ 
RRM & Radio resource management \\ 
RSRP & Reference signal received power \\ 
RW & Random-walk \\ 
SAC & Soft actor-critic \\ 
SON & Self-organizing network \\ 
TDM & Time-division multiplexing \\ 
UAV & Unmanned aerial vehicles\\ 
UE & User equipment \\
\hline
\end{tabular} 
\end{table}
\begin{table}[t!]
\centering
\caption{List of symbols and notations.}
\label{Notations}
\begin{tabular}{p{.15\columnwidth}|p{.75\columnwidth}}
\hline
\textbf{Symbol} & \textbf{Description} \\ 
\hline 
$\alpha$ & CQL hyperparameter \\ 

$\beta$ & Discount factor \\ 

$\pi_i(a^i \mid o^i )$ & Policy of agent $i$ \\ 

$a_i(t)$ & Action of agent $i$ at time step $t$ \\ 

$A(t)$ & Joint action space \\ 

$C_j(t)$ & Instantaneous rate of UE $j$ \\ 

$\Bar{C}_j$ & Average rate of UE $j$ \\ 

$\Bar{C}_{\text{sum}}$ & Sum rate \\ 

$\Bar{C}_{5 \%}$ & 5-percentile rate \\ 

$\Bar{C}_{\text{score}}$ & Score function \\ 

$d_0$ & Minimum distance between two APs \\ 

$d_1$ & Minimum distance between an AP and UE \\ 

$\mathcal{D}$ & Offline dataset \\ 

$h_{ij}$ & channel between UE $j$ and AP $i$ \\ 

$I$ & Number of APs \\ 

$J$ & Number of UEs\\ 

$L$ & Length of the network \\ 

$N$ & Number of top users \\ 

$o_i(t)$ & Local observations of agent $i$ at time step $t$ \\ 

$PL_{ij}$ & Path loss \\ 

$Q(s,a)$ & Q-function \\ 

$r(t)$ & Immediate reward at time step $t$\\ 

$S(t)$ & Overall state space \\ 

$w_j(t)$ & Weighting factor of user $j$ at time $t$\\

\hline
\end{tabular} 
\end{table}

\section{System Model}\label{sec:system_model}
\begin{figure}[t!]
    \centering    \includegraphics[width=1\columnwidth,trim={2.5cm 0 2.5cm 0},clip]{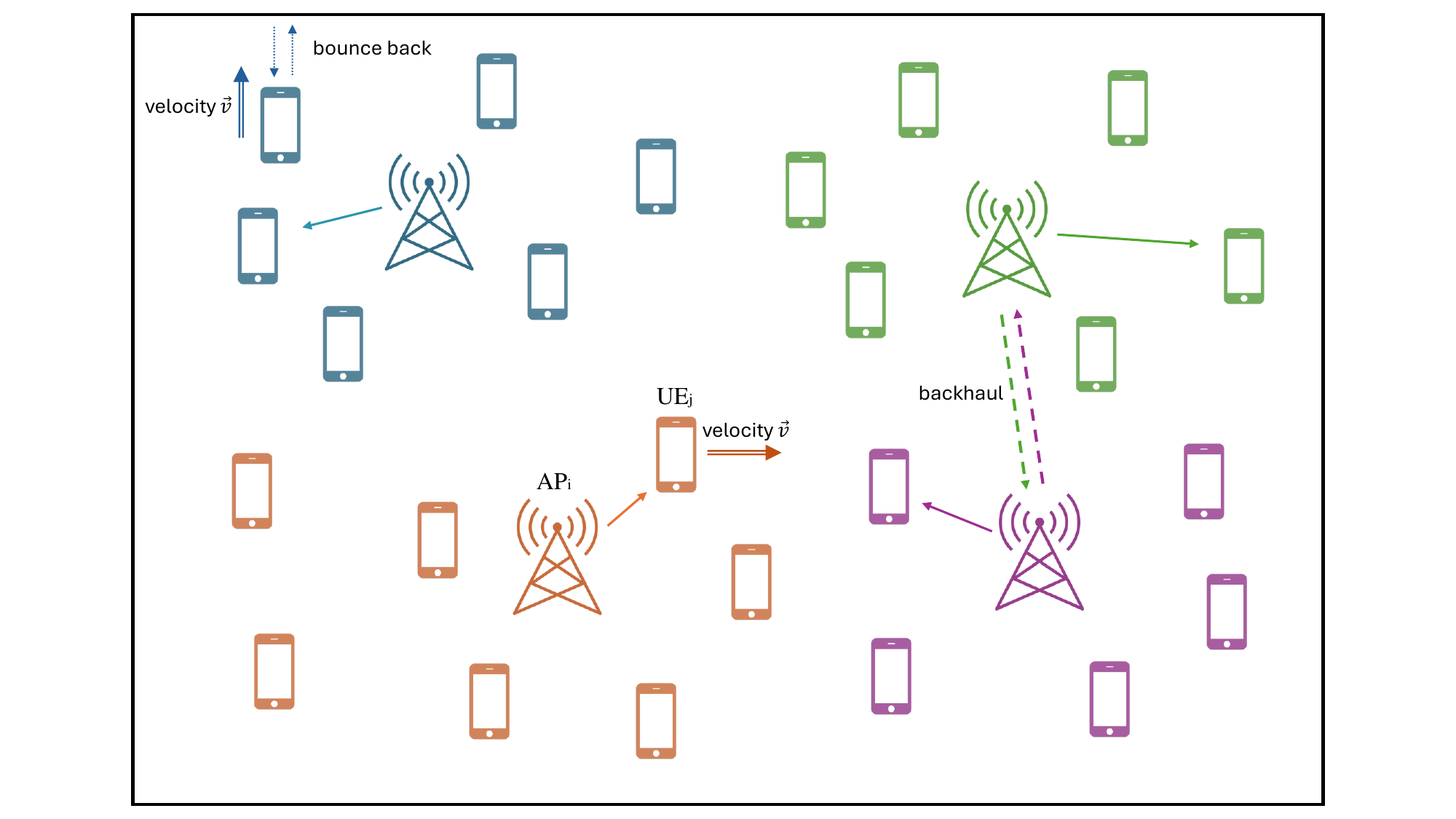} \vspace{2mm}
    \caption{A wireless environment consists of $I$ APs and $J$ UEs. Each UE is associated with only one $AP$, which chooses one of its associated UEs to serve at a time. 
    }
    \vspace{0mm}
    \label{sys_mod}
\end{figure}

Consider the downlink of a cellular system as illustrated in Fig.~\ref{sys_mod}. Assume an $L \times L$ square network, and consider $J$ user equipment (UEs) transmit their data to $I$ access points (APs) during $T$ discrete time intervals, forming an episode. At the beginning of each episode, APs and UEs are randomly deployed following a uniform distribution on the coordinates.
%
%
During each episode, the position of each AP is fixed, while UEs move randomly within the network's borders with a fixed velocity $v_t \in [0,1]$ m$/$s. To elaborate on a practical scenario, we set three thresholds:
\begin{enumerate}
    \item There is a minimum distance $d_0$ between any two APs
    \begin{align}
        d_{i^{\prime} i} > d_0, \quad &\forall \: i^{\prime},i \in {1, \cdots, I}, 
        \mathrm{~and~} i^{\prime} \neq i,
    \end{align}
    where we keep sampling new APs positions until meeting the threshold.
    
    \item There is a minimum distance $d_1$ between each AP and each UE
    \begin{align}
        d_{i j} > d_1, \quad &\forall \: i \in {1, \cdots, I}, 
        \mathrm{~and~}\forall \: j \in {1, \cdots, J},
    \end{align}
    where we keep sampling new UEs positions until meeting the threshold.
    
    \item All UEs are prohibited from moving outside the network's borders, where we consider a bounce back when a UE aims to move outside the borders.
\end{enumerate}

The received signal of UE $j$ from AP $i$ at time $t$ is
\begin{equation}
    \label{received_signal}
    y_j(t) = h_{ij}(t) x_i(t) + \sum_{i^{\prime} \neq i} h_{i^{\prime}j}(t) x_{i^{\prime}}(t) + n_j(t),
\end{equation}
where $n_j(t)$ is the additive white Gaussian noise (AWGN) whose power is $\sigma^2$ and the channel between AP $i$ and UE $j$ is denoted as $h_{ij}$ and comprises:
\begin{itemize}
    \item path-loss: we follow the 3GPP indoor model~\cite{3gpp_indoor}
    \begin{equation}
        PL_{ij} = 15.3 + 37.6 \log(d_{ij}) + 10,
    \end{equation}
    \item shadowing: we consider the log-normal effect with a standard deviation $\sigma_{sh}$~\cite{3gpp_indoor}, and
    \item short-term fading: we consider frequency-flat Rayleigh fading on all links in the network.
\end{itemize}

Each UE is associated with one of the APs at the beginning of each episode according to the reference signal received power (RSRP)~\cite{3gpp_measure}. In other words, a UE $j$ is associated with AP $i$ that records the max RSRP among all APs
\begin{equation}
    i = \argmax{i^{\prime}} \: \text{RSRP}_{i^{\prime}j}, \quad \forall \: i^{\prime} \in \{1, \cdots, I\}.
\end{equation}
%
%
Then, each time $t$, each user selects one of its associated UEs to serve. To this end, the instantaneous rate and SINR of UE $j$ that is associated with AP $i$ are, respectively,
\begin{align}
    \label{inst_rate}
    &C_j (t) = \log_2 \big(1 + \gamma_j(t) 
    \big), \\
    &\gamma_j(t) =  \frac{|h_{ij}(t)|^2 \: p_t}{ \sum_{i^{\prime} \neq i} |h_{i^{\prime}j}(t)|^2 \: p_t + \sigma^2},
\end{align}
where $p_t$ is the transmit power. 

Then, in an episode, the average rate of UE $j$, $\Bar{C}_j$, is
\begin{equation}
\label{avg_throughput}
    \Bar{C}_j = \frac{1}{T} \sum_{t=1}^T C_j (t).
\end{equation}
To simplify the notation, time-dependent variables will be referred to without explicitly indicating the time dependency $(t)$, assuming it remains implicit unless specified otherwise.

\subsection{Problem Formulation}
The main objective in RRM problems is to find the optimal serving policy for each AP that maximizes the average rate across all users. However, simply formulating the problem to maximize \eqref{avg_throughput} leads to a solution that invariably favors the user with the best SINR, thus disregarding fairness across users. To address this, the problem formulation should balance the sum and tail rate, ensuring a more equitable distribution of resources across all users.

Bearing this in mind, the sum rate is 
\begin{equation}
    \label{sum_rate}
    C_{\text{sum}} = \sum_{j=1}^J \Bar{C}_j,
\end{equation}
whereas the tail rate, \textit{i.e.}, the 5-percentile rate, is formulated as
\begin{equation}
    \label{5_perc_rate}
    C_{5 \%} = \text{max} \: \: C \: \: \text{s.t.} \: \: \mathbb{P} [\Bar{C}_j \geq C] \geq 0.95, \: \: \forall \: j \in \{1, \cdots, J\}.
\end{equation}
Next, we define the score function, $C_{\text{score}}$, as the linear combination of these rates
\begin{equation}
    \label{R_score}
    C_{\text{score}} = \mu_1 \: C_{\text{sum}} + \mu_2 \: C_{5 \%},
\end{equation}
where $\mu_1 \in \mathbb{R}_{+}$ and $\mu_2 \in \mathbb{R}_{+}$ are user chosen parameters. 

We are now ready to cast the RRM problem as
\begin{equation}
    \label{P1}
    \mathbf{P1:}\: \: \: \: \underset{A}{\max}  \: \: \sum_{t=1}^T C_{\text{score}}, 
\end{equation}
where ${A}$ (defined in Section~\ref{sec:MARL}) is the action space that describes the jointly serving policies of all APs.

Directly optimizing~\eqref{P1} imposes several challenges due to time dependency between actions that affect both sum and tail rates. In addition, the $5$-percentile rate is challenging to optimize due to its instability as it can not be formulated in a closed form as a function of the system parameters. 
Alternatively, the authors in~\cite{naderializadeh2021resource} proposed a more sophisticated approach to address this complex optimization. Consider the weighting factor $w_j(t)$ of user $j$ at time $t$, which can be recursively obtained as
\begin{align}
    \label{eq:wj}
    &w_j(t) = \frac{1}{\Tilde{C}_j(t)},\\ \nonumber
    &\Tilde{C}_j(t) = \eta \: C_j(t) + (1-\eta) \: \Tilde{C}_j(t-1), \\ \nonumber
    &\Tilde{C}_j(0) = C_j(0),
\end{align}
where $\eta$ is a running average parameter and $\Tilde{C}_j$ is the long-term average rate of user $j$ at time $t$. The proportional fairness (PF) ratio $\text{PF}_j$ of user $j$ at time $t$ is defined as the product of \eqref{eq:wj}, the weighting factor, and  \eqref{inst_rate}, the instantaneous rate, 
\begin{equation}
    \text{PF}_j = w_j \: C_j.
\end{equation}
The PF factor indicates that if a user has low rates for a long time, its PF factor increases subsequently. Optimizing the PF factor is easier and directly influences the objective function, maximizing the score function, $C_{\text{score}}$. The optimization problem is now formulated as
\begin{equation}
    \label{P1_edited}
    \mathbf{P1:}\: \: \: \: \underset{{A}}{\max}  \: \: \sum_{t=1}^T \sum_{j=1}^J (w_j)^{\lambda} \cdot C_j, 
\end{equation}
where $\lambda \in [0,1]$ controls the trade-off between the sum-rate and the $5$-percentile rate.
%
%

Following~\cite{naderializadeh2021resource} and to generalize the problem, we use the PF ratio to prioritize the associated UEs of each AP to limit the number of UEs that each AP can choose from to $N$ users at each time $t$. These $N$ UEs are the top $N$ in PF ratios among all the associated UEs to a specific AP. This step is common in unifying the action space size among network configurations.

\section{MARL Formulation}\label{sec:MARL} 
In this section, we formulate the problem using a partially observable Markov decision process (PO-MDP) and present an online solution using MARL.

\subsection{Partially-Observable Markov Decision Process}
To solve the optimization problem in~\eqref{P1}, we rely on MARL. In particular, we assume each AP is an individual agent contributing to his policy toward maximizing the score function. In PO-MDP, each agent $i$ observes his local observation ${o}_i$, takes an action ${a}_i$, and receives a reward $r$. Jointly, the local observations of all agents together form the state space, and the actions of all agents form the action space ${A}$. Sharing the local observations among all agents converts the problem into fully observable MDP. The PO-MDP formulation is detailed as follows:
\begin{enumerate}
    \item \textbf{Local Observations:} Each agent $i$ observes a tuple comprised of the SINR, $\gamma_j$, of its top $N$ users and their weighting factor $w_j$. Then, each AP has a local observation $o_i = \left( (\gamma_{i,1}, w_{i,1}), \cdots, (\gamma_{i,N}, w_{i,N}) \right)$ whose size is $2 \times N$. The state space is the concatenation of all local observations $S = (o_1, \cdots, o_I)$ whose size is $2 \times N \times I$, i.e., the local observations of all $I$ APs. 

    \item \textbf{Actions:} The action space of each agent $a_i$ at time $t$ comprises the UEs among its top $N$ users chosen to be served and additional silent action (the agent turned itself off). The size of the individual action space is $N + 1$. The joint action space $A = (a_1, \cdots, a_I)$ has a size of $(N+1) \times I$.

    \item \textbf{Rewards:} We use a joint centralized reward function based on the actions of all APs. The reward 
    is 
    \begin{equation}
        r = \sum_{j=1}^J w_j^{\lambda} \: C_j.
    \end{equation}

    \item \textbf{Policies:} Each agent's policy, denoted as $\pi_i(o_i|a_i)$, maps the chosen action at each visiting observation. The global policy of the environment $\pi(S|A)$ maps the joint action to the global state. The goal is to find the optimal global policy that maximizes the rewards.
\end{enumerate}

\subsection{Online RL}\label{ssec:online-rl-sac}
Online RL, especially deep RL, efficiently solves complex and large-scale problems, such as the presented RRM problem. In this subsection, we define the preliminaries of single-agent RL needed to better define the proposed offline MARL scheme.
%
In this work, we choose discrete SAC (an actor-critic algorithm for environments with discrete actions)~\cite{pmlr-v80-haarnoja18b} as our online RL algorithm due to its stability compared to DQN, which usually sticks to local minimums and saddle points. 
SAC is a model-free, off-policy RL algorithm that optimizes the current policy by utilizing experiences from previous visits (across various policies). It uniquely maximizes the policy's rewards and entropy, promoting continuous and random exploration of the environment. This dual objective ensures that SAC seeks optimal actions and maintains sufficient exploration to avoid local optima.
Actor-critic architectures rely on policy evaluation and policy improvement alternately~\cite{SAC_paper}.  
SAC computes the Q-function iteratively via the policy evaluation loss
\begin{align}
    \label{Eval_loss}
    \mathcal{L}_{\text{eval}} \!= & \hat{\mathbb{E}} \!\Bigg[\! \Big(r \!+\! \beta \: \hat{\mathbb{E}}_{a^{\prime}\sim\pi^k(a^{\prime} | s^{\prime})} \big[\hat{Q}^{(k)}(s^{\prime},a^{\prime}) 
   \!-\! Q(s,a)\big] \!\Big)^{\!2} \!\Bigg],
\end{align}
where $\hat{\mathbb{E}}[\cdot]$ is the empirical expectation over samples $(s,a,r,s^{\prime})$, $\hat{Q}^{(k)}$ is the current estimate of the Q-function $Q$ at iteration step $k$, $s^{\prime}$ is the next state, $a^{\prime}$ is the next action, and the Q-function $Q(s,a)$ is usually modeled as a neural network parameterized by weights $\theta$. The latter updates the policy towards maximizing the expected Q-function through the policy improvement loss
\begin{align}
    \label{Improv_loss}
    \mathcal{L}_{\text{imp}} = \:& - \hat{\mathbb{E}}_{a \sim \pi^k (s | a)} \Bigg[ \hat{Q}^{k} (s,a) - \log \: \pi (a|s) \Bigg],
\end{align}
where the term $\log \: \pi (a|s)$ is the entropy regularization parameter, and the policy $\pi (a | s)$ is usually modelled as a neural network parameterized by weights $\phi$.

\begin{figure*}[t!]
    \centering    \includegraphics[width=1.7\columnwidth,trim={0cm 0cm 0cm 0cm},clip]{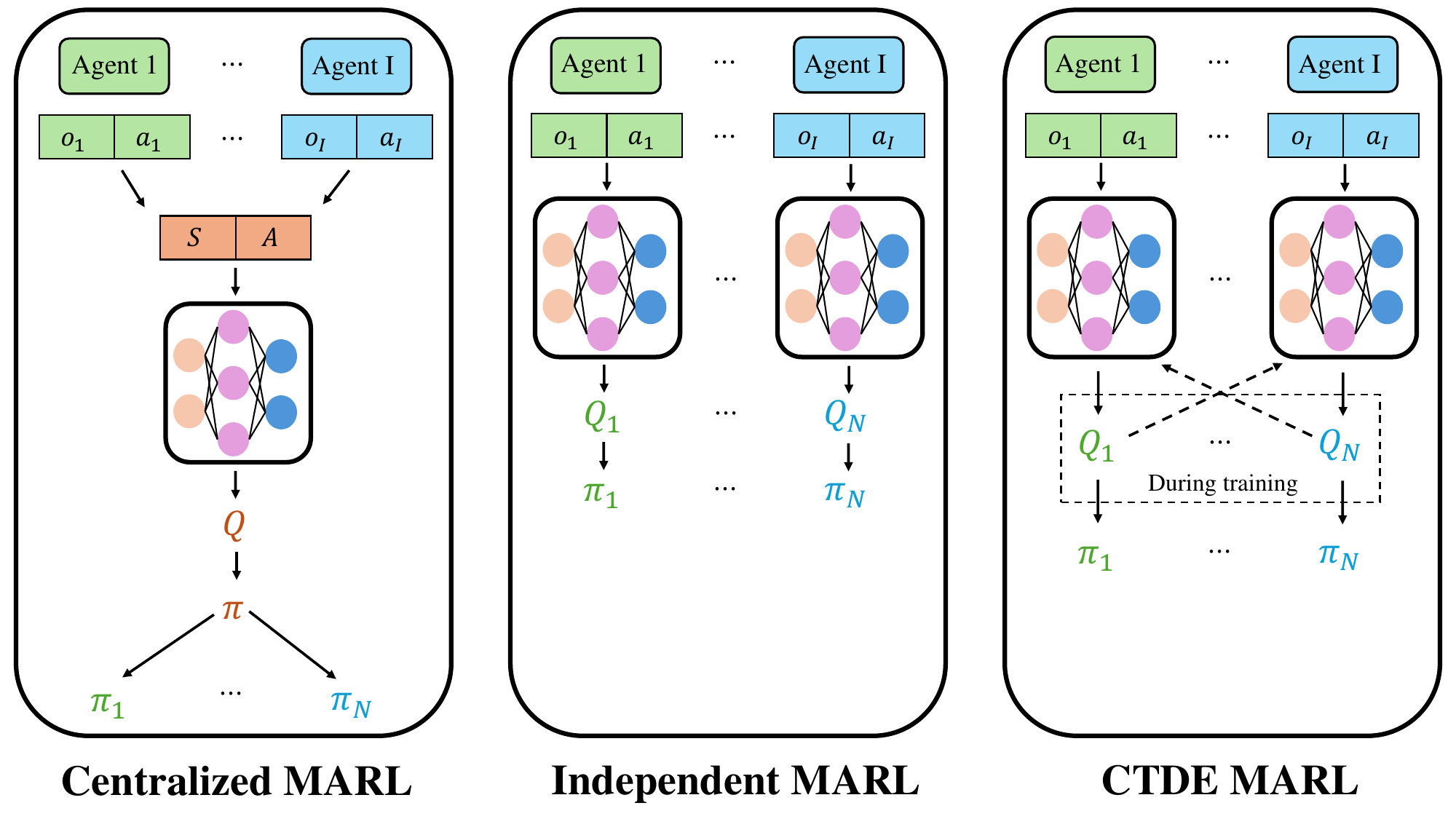} \vspace{2mm}
    \caption{An illustrative comparison between centralized MARL, independent MARL, and centralized training decentralized execution MARL. As shown, C-MARL models a joint Q-function using a single neural network, and the policies are drawn from the joint Q-function. In contrast, in I-MARL and CTDE-MARL, each agent models its Q-function as an independent neural network.}
    \vspace{0mm}
    \label{MARL_Variants}
\end{figure*}

\subsection{Online MARL}
Multiple agents cooperate toward the joint goal in the cooperative multi-agent RL setting. There are numerous variants of the MARL problem:

\subsubsection{Centralized MARL (C-MARL)} In this MARL setting, all agents are considered as one agent with one joint observation (state space) and one joint action space. The target is to find the optimal joint policy $\pi$ that maps the state space to the joint action space. Then, individual policies are extracted from the joint policy. Hence, the loss equations for the SAC algorithm in that case is
\begin{align}
\label{Eval_loss_centralized}
\mathcal{L}_{\text{eval}}^{\text{C}} \!=\:& \hat{\mathbb{E}} \!\Bigg[\!\Big(\!r \!+\! \beta \,\hat{\mathbb{E}}_{A^{\prime}\sim\pi^k(A^{\prime} | S^{\prime})} \hat{Q}^{(k)}(S^{\prime}\!,\!A^{\prime}) 
\!-\! Q(S,\!A)\!\Big)^{\!2} \!\Bigg],
\end{align}
where $\hat{Q}^{(k)}$ is the current estimate of the joint Q-function at iteration step $k$.
Similarly, the policy improvement loss for each agent is computed as follows:
\begin{align}
    \label{Improv_loss_centralized}
\mathcal{L}_{\text{imp}}^{\text{C}} = \:& - \hat{\mathbb{E}}_{A \sim \pi^k (A | S)} \Bigg[ \hat{Q}^{k} (S,A) - \log \: \pi (S|A) \Bigg].
\end{align}
C-MARL efficiently finds the optimal policies as it utilizes all agents' information jointly but at the cost of high complexity due to the vast state and action space dimensions.

\subsubsection{Independent Training MARL (I-MARL)} 
In I-MARL, each agent individually optimizes its policy using a distinct neural network that maps its observations to its action space. Consequently, each agent independently implements its own SAC algorithm, relying solely on its observations, actions, and accumulated experiences to refine its strategy.
Thus, the policy evaluation loss of each agent is computed as
\begin{align}
    \label{Eval_loss_agents}
    \mathcal{L}_{\text{eval}}^\text{I}\!= \,& \hat{\mathbb{E}} \!\Bigg[\! \Big(\!r \!+\! \beta \: \hat{\mathbb{E}}_{a_i^{\prime}\sim\pi_i^k(a_i^{\prime} | o_i^{\prime})} \!\hat{Q_i}^{\!(k)}\!(o_i^{\prime},\!a_i^{\prime}) 
   \!-\! Q_i(o_i,\!a_i) \!\Big)^{\!2} \!\Bigg], 
\end{align}
where $\hat{Q_i}^{(k)}$ is the current estimate of the Q-function of agent $i$ at iteration step $k$.
Similarly, the policy improvement loss for each agent is computed as follows:
\begin{align}
    \label{Improv_loss_agents}
    \mathcal{L}_{\text{imp}}^\text{I} = \:& - \hat{\mathbb{E}}_{a_i \sim \pi_i^k (o_i | a_i)} \Bigg[ \hat{Q_i}^{k} (o_i,a_i) - \log \: \pi_i (a_i|o_i) \Bigg].
\end{align}
I-MARL overcomes the complexity of C-MARL. However, it performs worse than C-MARL due to the missing information about the observations of other agents.
    
\subsubsection{Centralized training decentralized execution MARL (CTDE-MARL)} In the CTDE setting, value decomposition \cite{su2021value} approximates the global value function through the sum of the individual action-value functions of each agent, \textit{i.e.},
\begin{align}
\label{value_dec_eq}
   Q_{tot}(s) = \sum_{i=1}^I \tilde{Q}^i(o_i),
\end{align}
where $\tilde{Q}^i(o_i)$ represents the contribution of each agent to the global Q-function. In this framework, a joint policy evaluation loss is calculated based on the critical contributions of each agent, while each agent independently computes its policy improvement loss. The CTDE-MARL evaluation loss is
\begin{align}
    \label{Eval_loss_CTDE}
    \mathcal{L}_{\text{eval}}^{\text{CTDE}} = \:& \hat{\mathbb{E}} \Bigg[ \Big(r + \beta \: \hat{\mathbb{E}}_{a_i^{\prime}\sim\pi_i^k(a_i^{\prime} | o_i^{\prime})} \sum_{i=1}^I \hat{Q_i}^{(k)}(o_i^{\prime},a_i^{\prime}) \\ \nonumber 
   &\qquad- \sum_{i=1}^I \tilde{Q}_i(o_i,a_i) \Big)^2 \Bigg].
\end{align}
Then, the policy of each agent is obtained from the optimized function $\tilde{Q}_i(o_i,a_i)$ as
\begin{equation}
\label{Policy_opt_ind}
\pi_i(a_i|o_i) = \mathbbm{1}\left\{ a_i = \argmax{a_i} \tilde{Q}_i(o_i,a_i)\right\}.
\end{equation}
During execution, each agent uses his policy in a decentralized fashion. CTDE-MARL overcomes the complexity of C-MARL and the inefficiency of I-MARL.

Fig.~\ref{MARL_Variants} illustrates the three online MARL variants: C-MARL, I-MARL, and CTDE-MARL, graphically highlighting their similarities and differences. Note that each agent calculates its action-value functions in I-MARL and CTDE-MARL. However, CTDE-MARL improves upon I-MARL by aggregating the individual action-value functions to contribute to a global action-value function during training, as denoted in \eqref{value_dec_eq} and \eqref{Eval_loss_CTDE}. This approach enhances coordination among agents by aligning their objectives toward the global objective.

\section{Offline MARL}\label{sec:OMARL}
\begin{figure*}[t!]
    \centering
    \includegraphics[width=2.04\columnwidth,trim={0cm 4.7cm 0cm 4.7cm},clip]{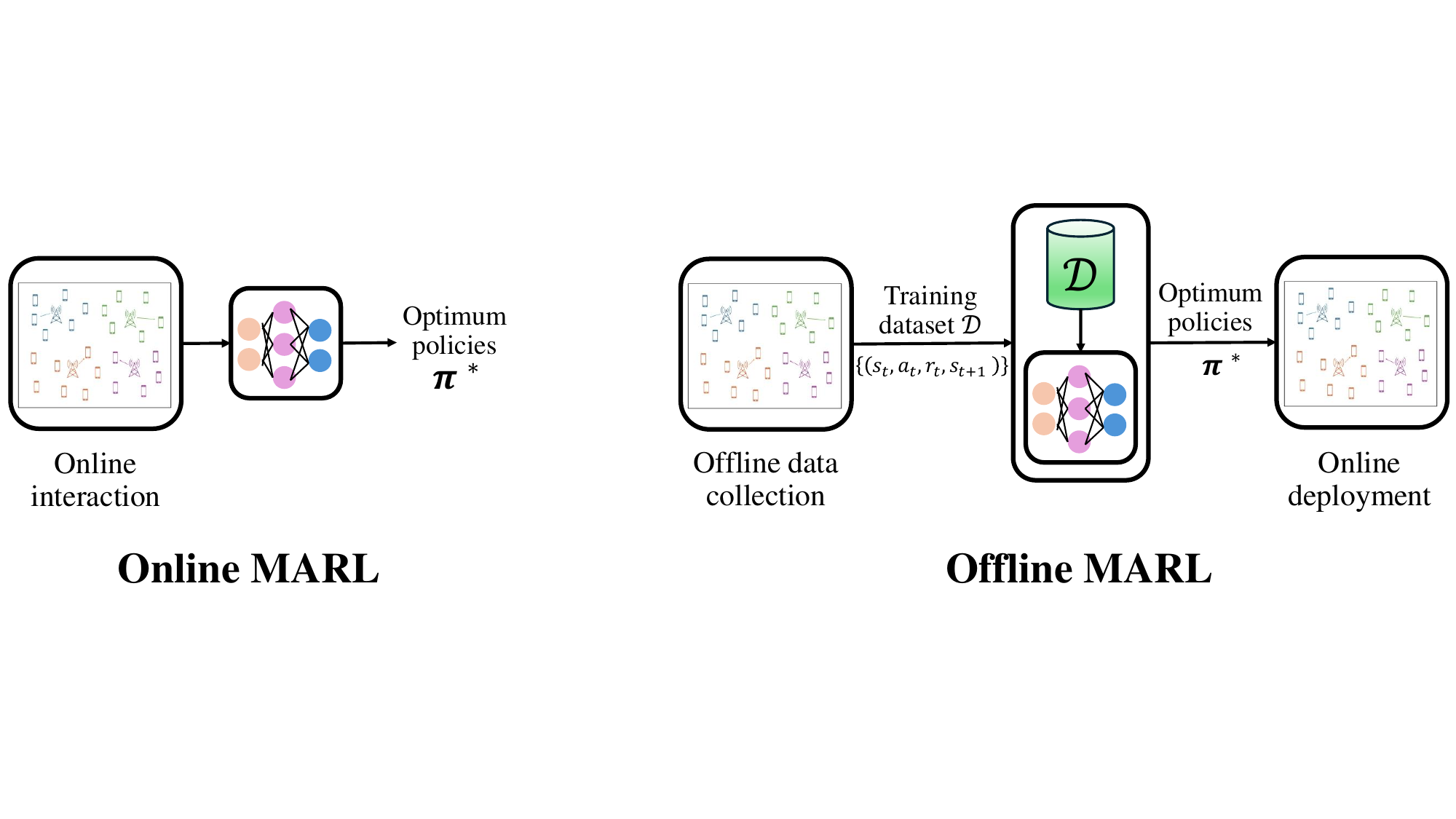} 
    \caption{An illustrative comparison between online MARL and offline MARL. Online MARL utilizes online interaction with the environment to optimize the policies. In contrast, offline MARL exploits offline datasets pre-collected using a behavioral policy. Offline MARL training uses the offline dataset, whereas optimum policies are used for online deployment.} 
    \vspace{-5mm}
    \label{Offline_MARL}
\end{figure*}
In the previous section, we presented online MARL and its variants. This section presents the proposed offline MARL scheme for the RRM problem. As shown in Fig.~\ref{Offline_MARL}, offline RL / MARL utilizes static offline dataset without any reliance on online interaction with the environment~\cite{levine2020offline}. This offline dataset is collected using behavioral policies from benchmark algorithms or random exploration. Since offline MARL only uses offline datasets, it removes the burden of online interaction in the RRM problem. For instance, to reach a sub-optimal policy in MARL algorithms, the agents need a large amount of online interaction with the environment. Therefore, the agents must visit as many state-action pairs as possible, which is costly regarding time and computations. Moreover, online MARL requires a high level of synchronization between the agents using a central unit that collects the agent's actions and re-distributes the calculated rewards. This creates a challenging communication overhead problem, which becomes even more cumbersome when information sharing is needed among the agents or centralized training.

Simply training the online SAC algorithm (described in Section \ref{ssec:online-rl-sac}) with an offline dataset usually fails.
This failure occurs because of the optimistic evaluation of the algorithm caused by the distributional shift between the learned and deployed actions, called out-of-distribution (OOD) actions.

However, recent advances in offline RL and offline MARL overcome this issue and have enabled deep RL and deep MARL algorithms to be used offline. For instance, behavior-constrained Q-learning (BCQ)~\cite{CQB_Paper} solved the OOD problem by limiting the distance between the selected actions for the policy to those in the dataset. In contrast, conservative Q-learning (CQL)~\cite{kumar2020conservative} adds a regularization term to the loss function to penalize large deviations between the selected actions and the actions in the dataset.

Next, we revise these algorithms in light of the RRM problem proposed in Section \ref{sec:system_model} and the SAC architecture discussed in Section \ref{sec:MARL}. 

{\LinesNumberedHidden
\begin{algorithm}[!t]
\SetAlgoLined

\textbf{Input:} Discount factor $\beta$, conservative penalty constant $\alpha$, number of agents $I$, number of training iterations $K$, number of gradient steps $G$, and offline dataset $\mathcal{D}$


Initialize network parameters

\For{\text{iteration} $k$ in $\{1$,...,$K$\}}{

\For{\text{gradient step} $g$ in $\{1$,...,$G$\}}{

Sample a batch $\mathcal{B}$ from the dataset $\mathcal{D}$

Estimate the C-MARL-CQL loss $\mathcal{L}_{\text{CQL}}^{\text{C}}$ in~\eqref{Eval_loss_centralized_CQL}

Estimate the policy improvement loss $\mathcal{L}_{\text{imp}}^{\text{C}}$ using~\eqref{Improv_loss_centralized}

Perform a stochastic gradient step based on the estimated losses.

}
}
\caption{Centralized multi-agent reinforcement learning using conservative Q-learning (C-MARL-CQL) algorithm.}
\label{Centralized_MARL_alg}  \vspace{0mm}
\end{algorithm}}

\subsection{Conservative Q-Learning}
In this work, we choose the CQL algorithm as the offline MARL approach for solving the RRM problem due to its robust performance on various offline RL / MARL problems. In addition, we introduce a new variation of the algorithm as we build the CQL algorithm on top of SAC architecture. As in the online case, next, we assess the algorithm for three variants: centralized (C), independent (I), and CTDE.

\subsubsection{C-MARL-CQL}
To implement the CQL algorithm in the C-MARL setting, which we refer to as C-MARL-CQL, the policy improvement loss is calculated as
\begin{align}
\label{Eval_loss_centralized_CQL}
\mathcal{L}_{\text{CQL}}^{\text{C}} \!=\! \frac{1}{2} \mathcal{L}_{\text{eval}}^{\text{C}} \!+\!\alpha \hat{\mathbb{E}}\! \bigg[\! \log \!\bigg(\!\!\sum_{A}
\exp \bigl( Q(S,A) \!\bigr)\!\!\bigg)\! 
    \!-\! \ Q(S,A)  \!\bigg],
\end{align}
where the term $\alpha \hat{\mathbb{E}} \Big[ \log \!\big(\!\sum_{A}
\exp \bigl( Q(S,A) \bigr)\!\big) - \ Q(S,A) \Big]$ is the regularization term (KL-divergence) and $\alpha > 0$ is a constant. Then, the policy improvement is performed using~\eqref{Improv_loss_centralized} as in the online case. The C-MARL-CQL procedure is detailed in  Algorithm~\ref{Centralized_MARL_alg}. 

{\LinesNumberedHidden
\begin{algorithm}[!t]
\SetAlgoLined

\textbf{Input:} Discount factor $\beta$, conservative penalty constant $\alpha$, number of agents $I$, number of training iterations $K$, number of gradient steps $G$, and offline dataset $\mathcal{D}$


Initialize networks parameters 

\For{\text{iteration} $k$ in $\{1$,...,$K$\}}{

\For{\text{gradient step} $g$ in $\{1$,...,$G$\}}{

Sample a batch $\mathcal{B}$ from the dataset $\mathcal{D}$

\For{\text{agent} $i$ in $\{1$,...,$I$\}}{

Estimate the I-MARL-CQL loss $\mathcal{L}_{\text{CQL}}^\text{I}$ using~\eqref{CQL_loss}

Estimate the policy improvement loss $\mathcal{L}_{\text{imp}}^\text{I}$ using~\eqref{Improv_loss_agents}

Perform a stochastic gradient step based on the estimated losses.

}
}
}
\caption{Independent multi-agent reinforcement learning using conservative Q-learning (I-MARL-CQL) algorithm.}
\label{Indendent_MARL_alg}  \vspace{0mm}
\end{algorithm}}

\subsubsection{I-MARL-CQL}
To implement the CQL algorithm in the I-MARL setting, named {I-MARL-CQL}, each agent $i$ computes its policy improvement loss as
\begin{align}
    \label{CQL_loss}
\mathcal{L}_{\text{CQL}}^\text{I}\!\!=\!\frac{1}{2} \mathcal{L}_{\text{eval}}^\text{I} \!+\! \alpha \hat{\mathbb{E}}\!\bigg[\!\!\log \!\bigg(\!\!\sum_{a_i}\!
\exp \bigl( Q_i(o_i,\!a_i) \!\bigr)\!\!\bigg)\! 
    \!-\! Q_i(o_i,a_i)  \!\bigg],
\end{align}
where the term $\alpha \hat{\mathbb{E}}\!\Big[\! \log\!\big(\!\sum_{a_i}
\exp \bigl( Q_i(o_i,a_i) \bigr)\!\big) \!-\! Q_i(o_i,a_i) \!\Big]$ is the regularization term for each agent. Then, the policy improvement is performed using~\eqref{Improv_loss_agents}, as in the online case.  Algorithm~\ref{Indendent_MARL_alg} details the I-MARL-CQL algorithm.

{\LinesNumberedHidden
\begin{algorithm}[!t]
\SetAlgoLined

\textbf{Input:} Discount factor $\beta$, conservative penalty constant $\alpha$, number of agents $I$, number of training iterations $K$, number of gradient steps $G$, and offline dataset $\mathcal{D}$


Initialize networks parameters 

\For{\text{iteration} $k$ in $\{1$,...,$K$\}}{

\For{\text{gradient step} $g$ in $\{1$,...,$G$\}}{

Sample a batch $\mathcal{B}$ from the dataset $\mathcal{D}$

Estimate the CTDE-MARL-CQL loss $\mathcal{L}_{\text{CQL}}^{\text{CTDE}}$ using~\eqref{Ind_CQL_CTDE}

\For{\text{agent} $i$ in $\{1$,...,$I$\}}{
Estimate the policy improvement loss $\mathcal{L}_{\text{imp}}^\text{I}$ using~\eqref{Improv_loss_agents}
}

Perform a stochastic gradient step based on the estimated losses.

}
}
\caption{Centralized training decentralized execution multi-agent reinforcement learning using conservative Q-learning (CTDE-MARL-CQL) algorithm.}
\label{CTDE_MARL_alg}  \vspace{0mm}
\end{algorithm}}

\subsubsection{CTDE-MARL-CQL}

Finally, the CQL loss in the CTDE form, which we call {CTDE-MARL-CQL}, is formulated as 
\begin{align}
\label{Ind_CQL_CTDE}
\mathcal{L}_{\text{CQL}}^{\text{CTDE}} \!&=\! \frac{1}{2} \mathcal{L}_{\text{eval}}^{\text{CTDE}} \\ 
&\quad\!+\!\alpha \hat{\mathbb{E}} \!\sum_{i=1}^I \! \bigg[\!\log \!\bigg(\! \sum_{a_i} 
\exp ( \tilde{Q}_i(o_i,a_i) )\bigg) \!-\! \tilde{Q}_i(o_i, a_i) \!\bigg]. \nonumber
\end{align}
The policy improvement is performed using~\eqref{Improv_loss_agents} for each agent. Lastly, the CTDE-MARL-CQL algorithm is detailed in Algorithm~\ref{CTDE_MARL_alg}.

We highlight that the key difference between each offline MARL scheme and its corresponding online MARL scheme is the carefully designed conservative term in the offline case, which pushes the learned policy close to the behavioral policy in the dataset. Therefore, Algorithms~\ref{Centralized_MARL_alg} to \textbf{\ref{CTDE_MARL_alg}} can be converted to the online counterpart by replacing in the appropriate policy evaluation loss function. Note that for the C-MARL-CQL, a single neural network is used to model the Q-function (\textit{i.e.}, the policy), whereas, in CTDE-MARL-CQL and I-MARL-CQL, each agent models its neural network.

\section{Numerical Results}\label{sec:results}
This section presents the numerical analysis of the designed offline MARL algorithms for the RRM problem. First, we present the implementation and the baseline models, then show the experimental results of the proposed model compared to the baseline models.

\subsection{Implementation and Baseline Models}

We consider a $100$ m $\times$ $100$ m square area with $I = 4$ APs (agents) and $J = 20$ UEs. At the beginning of each episode, one random environment is sampled with different AP positions and UEs' initial positions. Each episode consists of $200$ time steps. We use $2$ hidden layers in actor and critic with $256$ neurons each. All simulations are performed on a single NVIDIA Tesla V100 GPU using the Pytorch framework. Simulation parameters are shown in Table~\ref{RRM_Parameters}.
\begin{table}[h!]
\centering
\caption{Simulation parameters}
\label{RRM_Parameters}
\begin{tabular}{p{.25\columnwidth}p{.2\columnwidth}|p{.25\columnwidth}p{.2\columnwidth}}
\toprule
\textbf{Parameter} & \textbf{Value} & \textbf{Parameter} & \textbf{Value} \\ \midrule
$I$ & $4$ & $J$ & $20$ \\
$N$ & $3$ & $L$ & $100$ \\
$d_0$ & $10$ m & $d_1$ & $1$ m \\
$v(t)$ & $1$ m/s & $PL_o$ & $10$ dB \\
$p_t$ & $10$ dBm & $T$ & $200$ \\
$\mu_1$ & $\sfrac{1}{M}$& $\mu_2$ & $3$ \\
$\lambda$ & $0.8$ & $\beta$ & $0.99$ \\
$\alpha$ & $1$ & Replay memory & $10^5$ \\
Actor $lr$ & $1e-5$ & Critic $lr$ & $1e-4$ \\
Layers & $2$ & Neurons & $256$ \\
Optimizer & Adam & Activation & ReLu \\
\bottomrule
\end{tabular} 
\end{table}
First, we show the performance of online C-MARL (SAC) compared to the baseline models and the famous online C-MARL (DQN) as a learning-based baseline. Afterward, we compared the online performance to C-MARL-CQL (SAC). Then, we present the performance of the proposed offline MARL schemes, \textit{i.e.,} \emph{C-MARL-CQL (SAC)}, \emph{I-MARL-CQL (SAC)}, and \emph{CTDE-MARL-CQL (SAC)}. Using SAC as our deep RL framework, we always compare it to DQN. In our simulation, we use \emph{(SAC)} after the name of the algorithm to refer to a scheme built on top of SAC architecture and \emph{(DQN)} after the name of the algorithm to refer to a scheme built on top of the traditional DQN architecture. Finally, we show the effect of the quality and size of the dataset on the offline training.

We collect different offline datasets using the experience of an C-MARL (SAC) agent and different sub-optimal benchmarks, respectively. Since the performance of the offline MARL algorithms is sensitive to the quality of the offline dataset, we adopt the online centralized MARL approach to collect good-quality data points. In addition, we test the effect of the size of the dataset on the performance of offline MARL schemes by collecting datasets with different sizes.

Besides the three developed MARL algorithms, namely C-MARL-CQL, I-MARL-CQL, and CTDE-MARL-CQL, we show the performance of four benchmarks from the literature:
\begin{enumerate}
    \item \textbf{Random-walk (RW):} At each time step $t$, each AP chooses randomly to serve one of its top $N$ UEs.

    \item \textbf{Greedy:} In the greedy method, each agent serves the user with the highest SINR among its top $N$ users at each time step $t$.

    \item \textbf{Time-division multiplexing (TDM):} At each time step $t$, all UEs are served equally, where each AP serves the UEs in a round-robin manner.
    
    \item \textbf{Information-theoretic link scheduling (ITLinQ):} It was proved in~\cite{6875098}, that ITLinQ algorithm reaches a sub-optimal policy. At each time step $t$, each AP sorts its top $N$ UEs regarding their PF ratios. Then, each AP performs an interference tolerance check for each UE to ensure that the interference level is lower than a threshold $M \text{SNR}^{\eta}_{mn}$. This AP is turned off if no UEs have lower interference than the threshold.
\end{enumerate}
\begin{figure}[ht!]
    \centering
    \subfloat[Rsum\label{sum_online}]{\includegraphics[width=0.95\columnwidth]{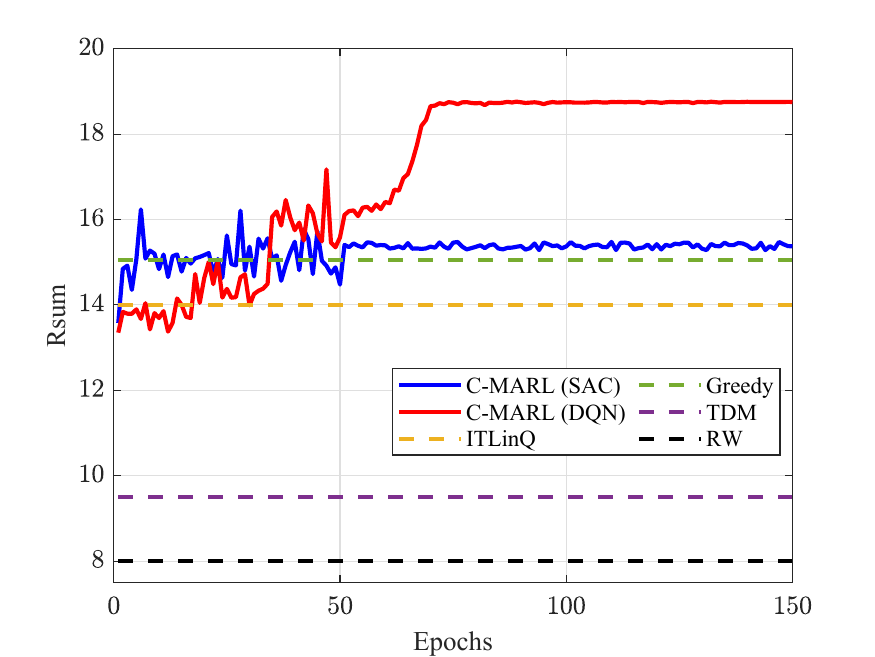}}\\
    \hskip -2.8ex
    \subfloat[Rperc\label{perc_online}]{\includegraphics[width=0.95\columnwidth]{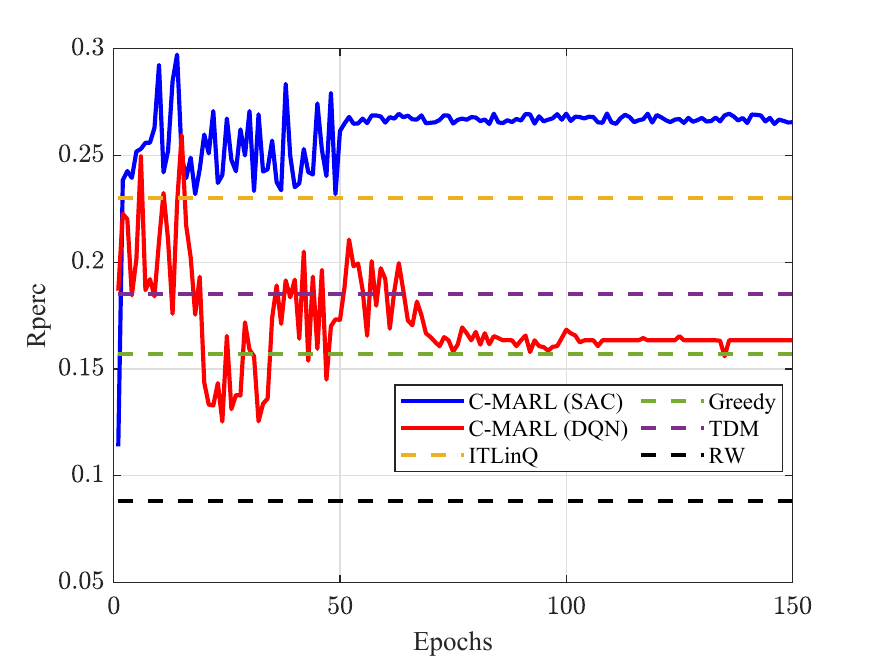}}\\
    \hskip -2.8ex
    \subfloat[Rscore\label{score_online}]{\includegraphics[width=0.95\columnwidth]{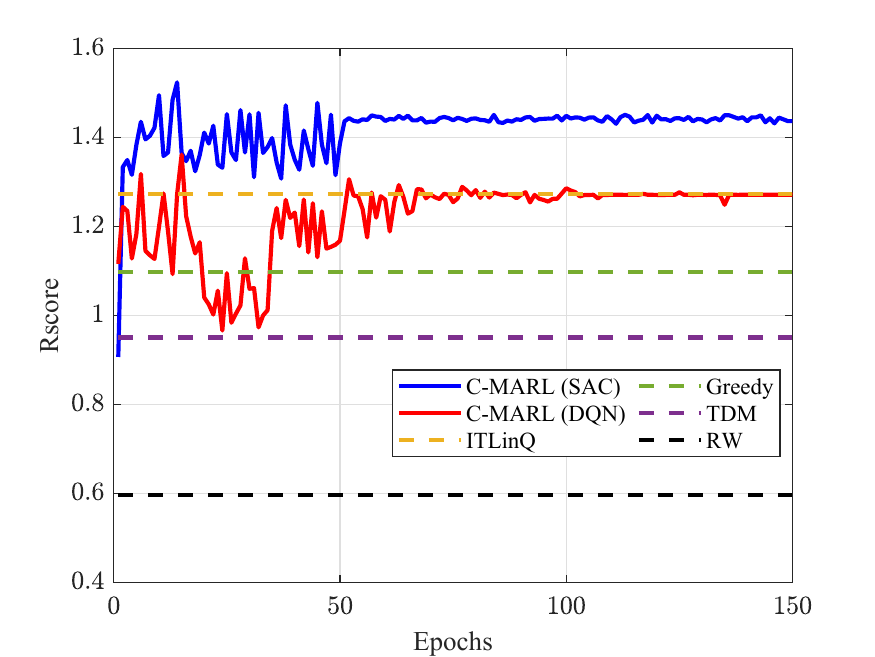}}
    \caption{The sum rate, $5$-percentile rate, and Rscore reported for C-MARL algorithm built on top of both SAC and DQN compared to other benchmark schemes.}
    \label{Rates_online} 
\end{figure}
\begin{figure}[ht!]
    \centering
    \subfloat[Rsum\label{sum_centr}]{\includegraphics[width=0.95\columnwidth]{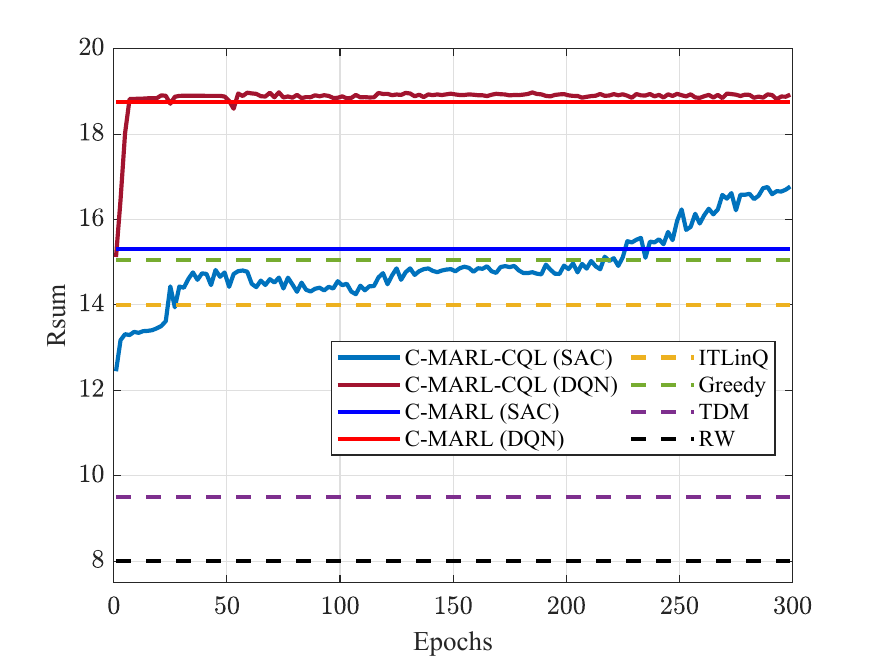}}\\
    \hskip -2.8ex
    \subfloat[Rperc\label{perc_centr}]{\includegraphics[width=0.95\columnwidth]{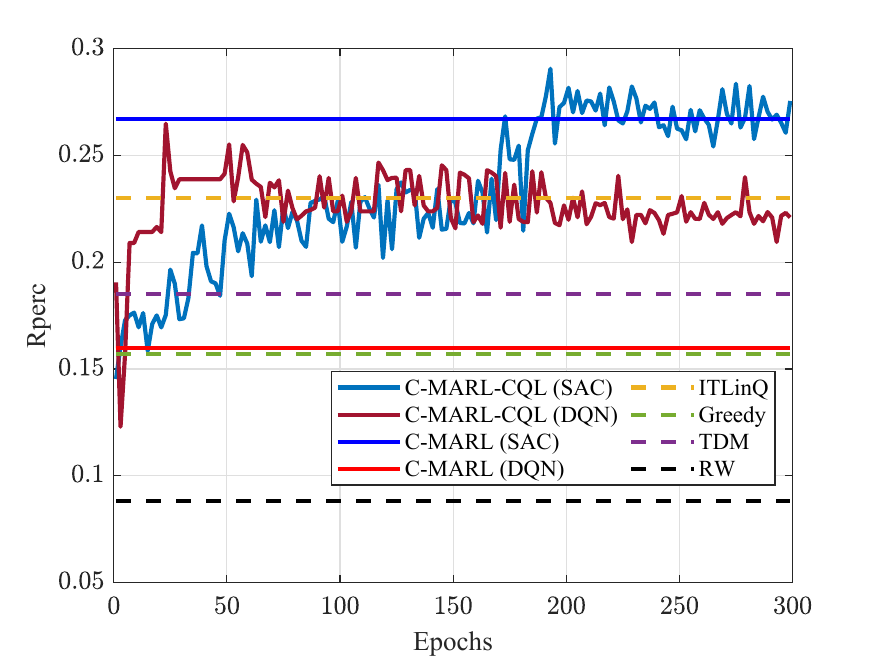}}\\
    \hskip -2.8ex
    \subfloat[Rscore\label{score_centr}]{\includegraphics[width=0.95\columnwidth]{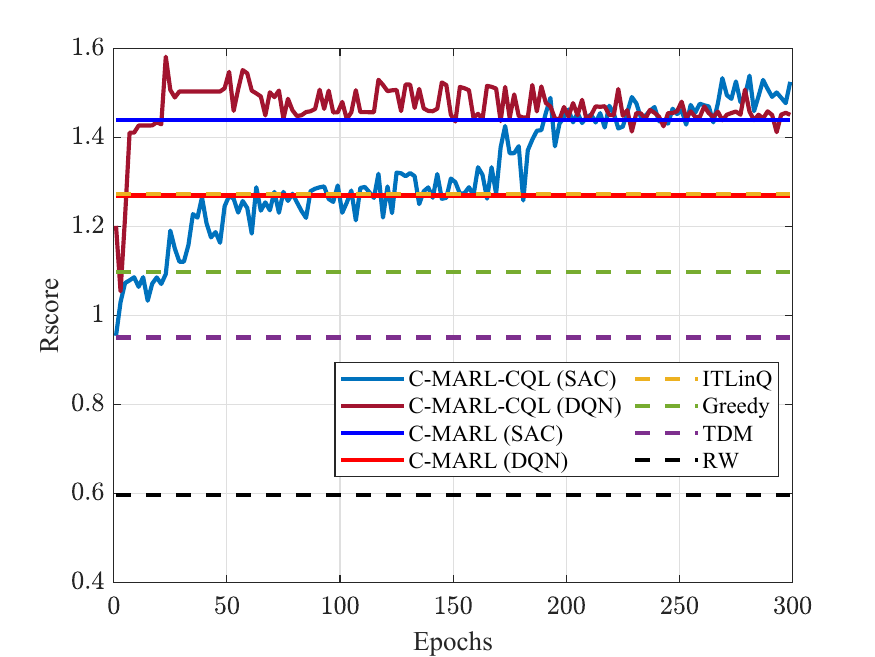}}
    \caption{The sum rate, $5$-percentile rate, and Rscore reported for the proposed C-MARL-CQL algorithm built on top of SAC and DQN compared to C-MARL and other benchmark schemes.}
    \label{Rates_centralized} 
\end{figure}


\subsection{Online Training and Dataset Collection}

In Fig.~\ref{Rates_online}, we show the training performance of a C-MARL (SAC) agent compared to the baseline schemes. In addition, we compare the developed C-MARL (SAC) scheme to a famous online RL algorithm, C-MARL (DQN) scheme, as a learning-based benchmark. First, the RW has the worst sum rate, $5$-percentile rate, and Rscore. The greedy algorithm maximizes the sum rate at the expense of the $5$-percentile rate. In contrast, the TDM scheme prioritizes maximizing the $5$-percentile rate over the sum rate. The ITLinQ benchmark has the highest Rscore among other baselines. The two online RL schemes, namely, C-MARL (SAC) and C-MARL (DQN), have the highest Rscore compared to all the baselines. We can notice in Fig.~\ref{sum_online} that the C-MARL (DQN) agent scores the largest sum rate, whereas the C-MARL (SAC) agent maintains a relatively good sum rate compared to other benchmarks. In contrast, the $5$-percentile rate of the C-MARL (DQN) agent drops noticeably compared to C-MARL (SAC), whose $5$-percentile rate is around $0.28$, as shown in Fig.~\ref{perc_online}. As a result, C-MARL (SAC) has a better overall Rscore than C-MARL (DQN), as in Fig.~\ref{score_online}.

\subsection{Offline Centralized Training}
In the next experiment in Fig.~\ref{Rates_centralized}, we show the performance of the proposed C-MARL-CQL (SAC) algorithm in terms of the sum rate, $5$-percentile rate, and Rscore. We compare the proposed algorithm to the traditional C-MARL-CQL (DQN), C-MARL (SAC), C-MARL (DQN) and other baselines as benchmarks to better evaluate it. In this experiment, we use the dataset, which is $16000$ data points, collected from an online SAC agent. In addition, we perform centralized training, \emph{i.e.}, C-MARL-CQL. As shown in Fig.~\ref{sum_centr} and Fig.~\ref{perc_centr}, and similar to the online case, the C-MARL-CQL (DQN) algorithm prefers to maximize the sum rate over the $5$-percentile rate, where the proposed C-MARL-CQL (SAC) algorithm sacrifices the sum rate to enhance the $5$-percentile rate. In Fig.~\ref{score_centr}, the C-MARL-CQL (SAC) algorithm outperforms the C-MARL-CQL (DQN) scheme, surpassing other baselines, including online benchmarks.
\begin{figure}[h!]
    \centering
    \subfloat[Rsum\label{sum_MARL}]{\includegraphics[width=0.95\columnwidth]{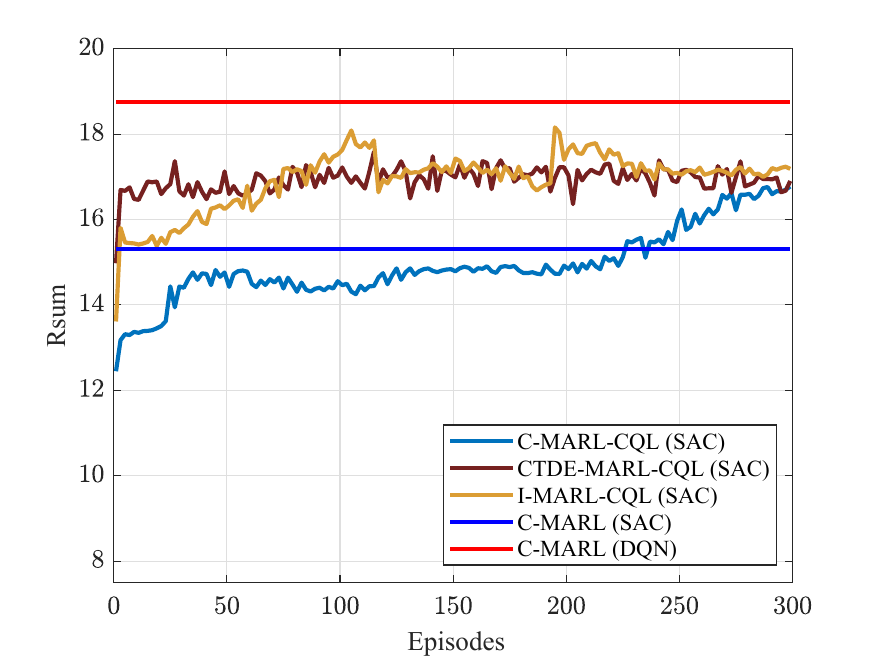}}\\
    \hskip -2.8ex
    \subfloat[Rperc\label{perc_MARL}]{\includegraphics[width=0.95\columnwidth]{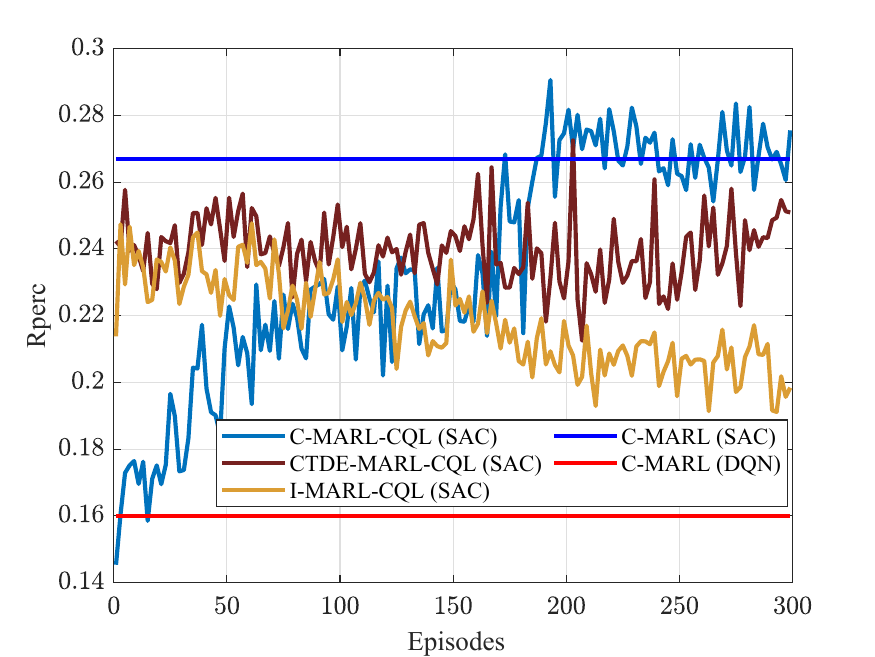}}\\
    \hskip -2.8ex
    \subfloat[Rscore\label{score_MARL}]{\includegraphics[width=0.95\columnwidth]{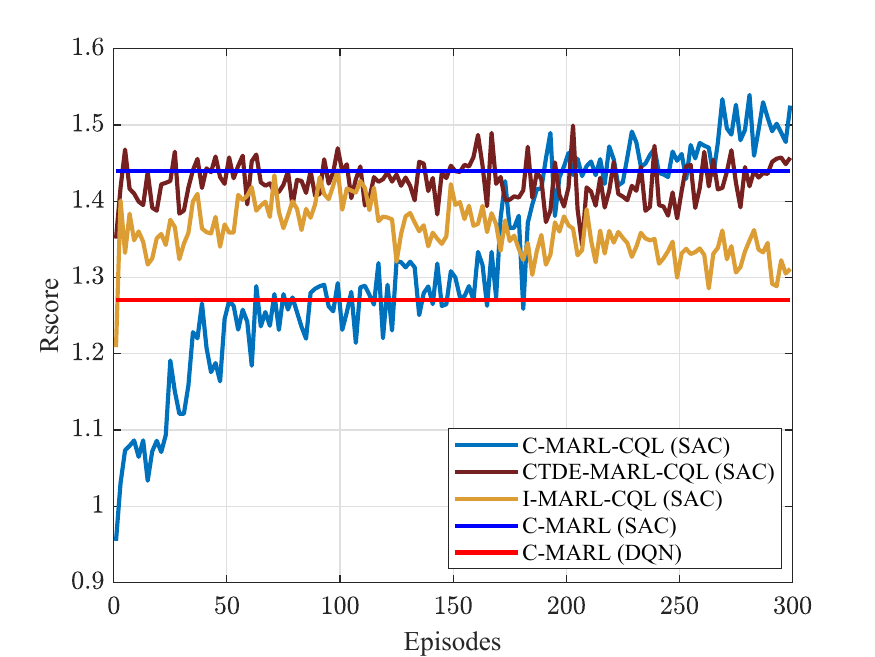}}
    \caption{The sum rate, $5$-percentile rate, and Rscore reported for the proposed C-MARL-CQL, I-MARL-CQL and CTDE-MARL-CQL built on top of SAC architecture.}
    \label{Rates_MARL} 
\end{figure}
\begin{figure}[!h]
    \centering
    \subfloat[Dataset Type\label{Rscore_dataset_type}]{\includegraphics[width=0.95\columnwidth]{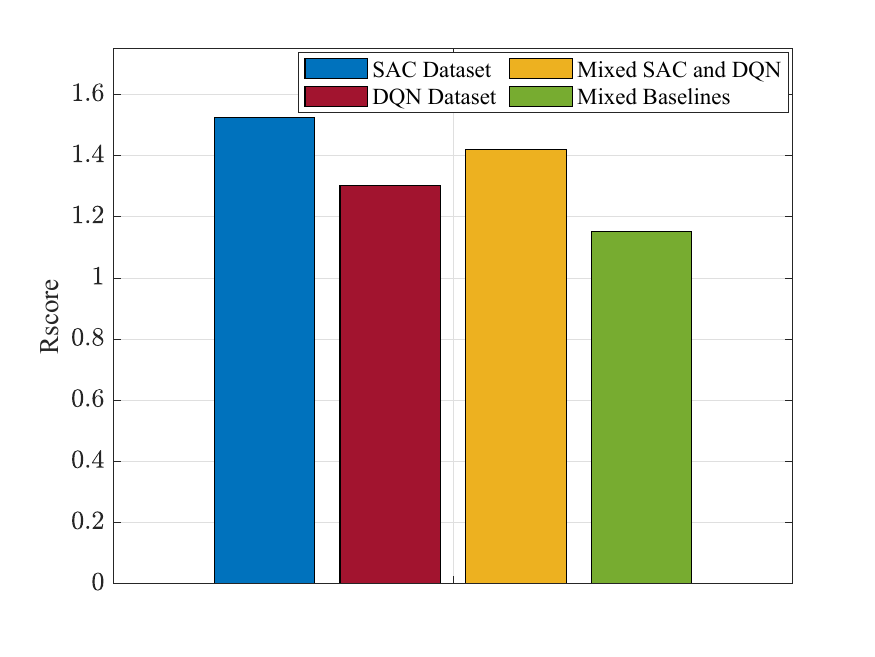}}\\
    \hskip -2.8ex
    \subfloat[Dataset Size\label{Rscore_dataset_size}]{\includegraphics[width=0.95\columnwidth]{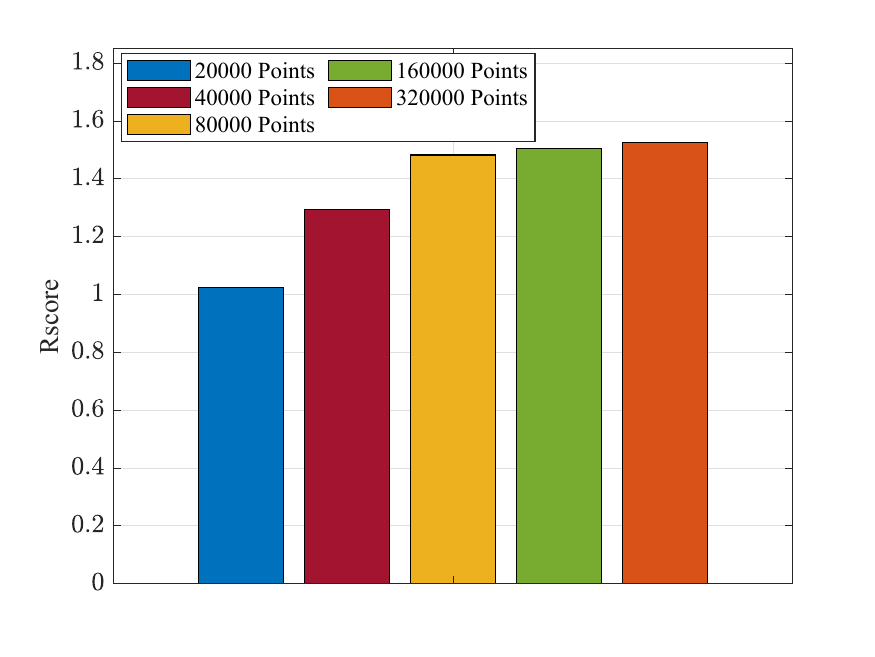}}
    \caption{The effect of the dataset on the overall performance of the proposed CTDE-MARL-CQL (SAC) scheme in terms of the achieved Rscore. Shown in (a) the effect of the quality of the collected dataset and (b) the effect of the dataset size.}
    \label{Datasets} 
\end{figure}

\subsection{Offline MARL Schemes}
In Fig.~\ref{Rates_MARL}, we report the rates of the proposed offline MARL schemes, namely, C-MARL-CQL, I-MARL-CQL and CTDE-MARL-CQL built on top of SAC architecture, compared to C-MARL (SAC) and C-MARL (DQN) as two benchmark schemes. We construct the three schemes on top of SAC architecture due to its stable and high rate convergence. As in Fig.~\ref{sum_MARL}, the three MARL schemes almost achieve the same sum rate, outperforming C-MARL (SAC). In contrast, C-MARL-CQL (SAC) has the highest $5$-percentile rate, relying on the availability of complete observations of all agents to find the optimum policies. Comparing CTDE-MARL-CQL (SAC) to I-MARL-CQL (SAC), we observe that CTDE-MARL-CQL (SAC), due to value function sharing among agents, approaches the $5$-percentile rate of C-MARL-CQL (SAC) with lower complexity, especially in execution. Hence, CTDE-MARL-CQL (SAC) outperforms C-MARL (SAC) Rscore with a tiny gap with C-MARL-CQL (SAC), as shown in Fig.~\ref{score_MARL}. This highlights the ability of the CTDE framework to overcome the complexity of centralized training and the poor performance of independent training.

\subsection{Dataset Quality}
Finally, we show the effect of the quality of the dataset and its size on the Rscore performance of the proposed CTDE-MARL-CQL (SAC) presented in Fig.~\ref{Datasets}. In particular, Fig.~\ref{Rscore_dataset_type} compares four sources of the offline dataset, \emph{i.e.}, online SAC, online DQN, the mixture of online SAC and online DQN, and the mix of other baselines\footnote{We only include experiences from RW, greedy, TDM, and ITLinQ benchmarks in this dataset.}. The quality of the policy used to collect the offline dataset directly reflects the achieved Rscore. A dataset collected from a good policy, such as online SAC, outperforms other datasets collected from online DQN agents and baseline policies. A mixture of SAC and DQN agents achieves a high score. This highlights that a mix of good and bad quality datasets can still be used to find a suitable policy offline~\cite{10529190}.

Similar to Fig.~\ref{Rscore_dataset_type}, Fig.~\ref{Rscore_dataset_size} shows the effect of the size of the dataset on the convergence of the Rscore of the proposed CTDE-MARL-CQL (SAC) algorithm. When using a small dataset of $20000$ points, the Rscore drops to $1$, similar to the performance of TDM. The lack of enough experience creates optimistic uncertainty in the CQL algorithm, forcing itself to converge to a saddle sub-optimal policy. When we increase the size of the dataset, the Rscore rapidly increases. Datasets with dimensions larger than $320000$ data points influence the convergence stability without achieving higher Rscore values.

\section{Conclusions}\label{sec:conclusions} 
%

This paper presents an offline MARL framework based on the SAC architecture and the CQL algorithm for optimizing resource management in wireless networks with multiple APs serving UEs. The framework introduces three variants: C-MARL-CQL (centralized training), I-MARL-CQL (independent training), and CTDE-MARL-CQL, tailored to balance computational complexity and policy performance. Numerical results demonstrate that the offline MARL framework significantly outperforms baselines, including random-walk, greedy algorithms, TDM, and ITLinQ, regarding the Rscore metric. Among the variants, CTDE-MARL-CQL achieves the best trade-off, offering reduced computational complexity compared to C-MARL-CQL while surpassing I-MARL-CQL in policy effectiveness.
Our analysis also underscores the importance of dataset quality and size in determining algorithm convergence and performance. High-quality behavioral datasets enhance rate optimization, while larger datasets contribute to stable convergence. These insights provide valuable guidance for offline MARL applications in wireless systems.

Future research will focus on extending this work to meta-offline RL and MARL, enabling dynamic adaptability to evolving environments and objectives. This direction can further enhance the scalability and robustness of offline MARL solutions in complex wireless communication scenarios.


\bibliographystyle{IEEEtran}
\bibliography{IEEEabrv,references}
\end{document}